\def\kms{\relax \ifmmode {\,\rm km\,s}^{-1}\else \,km\,s$^{-1}$\fi}    
   
\def\farcs{\hbox{$.\!\!^{\prime\prime}$}}   
\def\arcdeg{\hbox{$^\circ$}}   
\def\arcmin{\hbox{$^\prime$}}   
   
\def\secd#1.#2{ #1\farcs#2 }               

\def\mincir{\ \raise-2.truept\hbox{\rlap{\hbox{$\sim$}}\raise5.truept   
    \hbox{$<$}\ }}   
\def\magcir{\ \raise-2.truept\hbox{\rlap{\hbox{$\sim$}}\raise5.truept   
    \hbox{$>$}\ }}

\def\nii{[N~{\sc ii}]}   
\def\oiii{[O~{\sc iii}]}   
 
\def\oii{[O~{\sc ii}]}   
\def\ha{H$\alpha$}  
\def\hb{H$\beta$}

\def\ariii{Ar~{\sc iii}}   
\def\ariv{Ar~{\sc iv}}   
\def\hii{H~{\sc ii}}   
\def\heii{He~{\sc ii}} 
\def\he{He} 
\def\hei{He~{\sc i}}  
\def\sii{[S~{\sc ii}]}   
\def\siii{[S~{\sc iii}]}   
\def\cbeta{$c_{\beta}$}

\documentclass[referee]{aa} 
\usepackage[dvips]{graphics}   
\begin{document}   
\title{Spectroscopy of planetary nebulae in M~33\thanks{Based on  
observations obtained at the 4.2m~WHT telescope   
operated on the island of La Palma by the Isaac Newton Group in the Spanish   
Observatorio del Roque de Los Muchachos of the Instituto de Astrofisica de   
Canarias.}}   
  
\author{  
L. Magrini     \inst{1},  
M.   Perinotto \inst{1},   
R.L.M. Corradi \inst{2},  
A.   Mampaso   \inst{3},
 }   
\offprints{L. Magrini\\    
e-mail: laura@arcetri.astro.it}   
\institute{  
Dipartimento di Astronomia e Scienza dello Spazio, Universit\'a di  
Firenze, L.go E. Fermi 2, 50125 Firenze, Italy  
\and   
Isaac Newton Group of Telescopes, Apartado de Correos 321, 38700 Santa    
Cruz de La Palma, Canarias, Spain   
\and    
Instituto de Astrof\'{\i}sica de Canarias, c. V\'{\i}a L\'actea s/n,  
38200, La Laguna, Tenerife, Canarias, Spain 
}  
  
\date{Received date 7 August 2002/Accepted date 29 November 2002}   
  
\abstract{ 
Spectroscopic observations of 48 emission-line objects of M~33 have
been obtained with the multi-object, wide field, fibre spectrograph
AF2/WYFFOS at the 4.2m~WHT telescope (La Palma, Spain).  Line
intensities and logarithmic extinction, \cbeta, are presented for 42
objects.  Their location in the Sabbadin \& D'Odorico diagnostic
diagram (\ha/\sii\ vs \ha/\nii) suggests that $>$70$\%$ of the candidates
are Planetary Nebulae (PNe). Chemical abundances and nebular physical
parameters have been derived for the three of the six  PNe where the 4363\AA\,
\oiii\ emission line was measurable.  These are disc PNe, located
within a galactocentric distance of 4.1~kpc, and, to date, they are
the farthest PNe with a direct chemical abundance determination.  No
discrepancy in the Helium, Oxygen and Argon abundances has been found
in comparison with corresponding abundances of PNe in our Galaxy.
Only a lower limit to the sulphur abundance has been obtained since we
could not detect any \siii\ line.  N/H appears to be lower than the
Galactic value; some possible explanations for this under-abundance 
are discussed.
} 
\authorrunning{Magrini et al.}    
\titlerunning{Spectroscopy of planetary nebulae in M~33}     
\maketitle    
   
\section{Introduction}

Chemistry of galaxies has been studied for a long time using
spectroscopy of the integrated stellar light. There are, however,
limitations to this method, such as the large abundance variations
from star to star and the consequent difficulty in the interpretation
of the data (McWilliam \cite{mcwilliam}), or the impossibility of using 
this technique in poorly populated regions, as in the haloes of
elliptical galaxies.\\
Planetary nebulae (PNe) with their prominent emission lines 
can be used to measure the metallicity  
in nearby galaxies.
In spite of the importance of the topic, few spectroscopic data have
been obtained for PNe in external galaxies, except for the Magellanic
Clouds.  PNe in the LMC and the SMC have been extensively studied by
several authors, for instance Dopita et al. (\cite{dopita}), Reyes et
al. (\cite{reyes}), and Richer ({\cite{richer}).  To date, there are
only few other galaxies.  In fact, abundance analysis
requires the determination of electron temperature and density, as
well as the intensities of various lines of different ionization
states.  The most important line for temperature determination, \oiii\
4363\AA\ is, for instance, from 50 to 200 times weaker than \oiii\
5007\AA\ (e.g. Ford et al. \cite{ford}). Before our work, the
farthest galaxy with direct abundance measurements in PNe was M~31
(750 kpc, Freedman et al. \cite{freedman}).  Chemical abundances of
PNe in M~31 have been investigated by  Jacoby \& Ciardullo
(\cite{jacoby1999}, hereafter JC99) and Stasinska et
al. (\cite{stas1998}), as well as by Hyung et al. ({\cite{hyung}) who
used the data from the previous work of Stasinska and collaborators.
A total of 45 PNe have been studied.  In the companion galaxies of
M~31, spectrophotometric data have been obtained for PNe
of NGC~185, NGC~205 (Richer \& McCall \cite{richer95}) and of M~32
(Stasinska et al. \cite{stas1998}; Hyung et al. {\cite{hyung}).  In
the Fornax galaxy, one PN is known and its chemical abundances have
been derived by Danziger et al. (\cite{danziger}) and Walsh et
al. (\cite{walsh97}).  Spectrophotometry of the two PNe in the
Sagittarius dwarf elliptical galaxy was done by Walsh et
al. (\cite{walsh97}).  Chemical abundances of PNe in NGC~6822 were
computed by Richer \& McCall (\cite{richer95}).  
Spectra of 5 PNe in the giant elliptical galaxy NGC~5128 
(Centaurus~A) at the distance of 3.5 Mpc were analyzed by Walsh et 
al. (\cite{walsh99}). These are the farthest PNe where chemical abundances 
have been measured. This was done using diagnostic line ratios of various ions 
without, however, a direct determination of the 
electron temperature. 

 The galaxy M~33 (NGC~598) is the third-brightest member of the Local
Group.  Its large angular size (optical size 53'$\times$83', Holmberg
\cite{holmberg}) and its intermediate inclination $i$=56\arcdeg\,
(Zaritsky \cite{zaritsky}) make it particularly suitable for studies
of spiral structure and stellar content (van den Bergh
\cite{bergh2000}).  A total of 131 candidate PNe are known in M~33
distributed throughout the whole galaxy (Magrini et al. \cite{magrini00};
\cite{magrini01}). The objects were identified with the following two criteria:
{\it i)} they should appear both in the \oiii\ and \ha+\nii\ images
but not in the continuum frame, and {\it ii)} they should have a
stellar point spread function.  In this paper, we present spectroscopy of 39 PN
 candidates and of 9 unclassified emission-line objects discovered in our previous
survey (Magrini et al. \cite{magrini00}, hereafter M00).  In Section~2
we describe the observations, the reduction procedures and the flux 
measurement with their de-reddening.  The analysis of the spectra and
the determination of the physical and chemical properties of the PNe
are presented in Section~3. A critical discussion on the chemical
abundances is given in Section~4.  Conclusions follow in Section~5.
 
\section{Observations and data analysis} 
 
Thirty-nine candidate PNe and nine unclassified objects with emission
lines and a non-negligible continuum, selected from the list of M00,
were observed on October 16-17, 2001.  We have used AF2/WYFFOS, the
multi-object, wide field, fibre spectrograph working at the prime
focus of the 4.2~m William Herschel Telescope (La Palma, Spain).  The
WYFFOS spectrograph was used with a single setup: the R600B grating
(600 line mm$^{-1}$) providing a dispersion of 3.0 \AA/pixel.  The
resulting spectral range, from 4300\AA\, to 7380 \AA, included the
basic lines needed for the classification of the objects as PNe and
the determination of their chemical abundances.  However, due to the 
arrangement of fibres at the spectrograph entrance, not all the
spectra start at the same wavelength and for 15 PNe the
$\lambda$4363\AA\ spectral region was not observed. The spectrograph
WYFFOS was equipped with a 1024$\times$1024 TEK CCD.  We used the
Small Fibre module which is made of 150 science fibres with 1.6 arcsec
diameter (90 $\mu$m) projected on the sky.
 
Our targets, their astrometry and their absolute \oiii\ 5007\AA\ and
\ha\ + \nii\ fluxes came from our previous INT+WFC images (M00, Magrini et
al. \cite{magrini01}).  The accuracy in their positions is better than
0.5 arcsec r.m.s..  Our set of targets was chosen among the PNe
spanning a range of 3 mag in \oiii\ $\lambda$5007 \AA\, from the
bright cutoff of their luminosity function, and included also several
of the unclassified objects of our sample, i.e emission-line objects
with a non-negligible continuum emission (M00).  Targets are
distributed rather uniformly over the face of the galaxy (Figure 1).
We used the remaining fibres to take simultaneous sky/background
spectra.  A total of 13 science exposures of 2400~s each were taken (5
in the first night and 8 in the second night) through light cirrus
with 1.5 arcsec seeing. \\ Several offset sky exposures using the same
fibre configuration were taken before and after the M~33 observations
in order to do a correct sky subtraction.
 
\begin{figure*}  
\label{position} 
\caption{Target positions in the M~33 galaxy. The i.d. numbers are  
from M00. The emission-line + continuum objects are shown with their i.d. 
number, preceded by a capital C.  The image is from the ESO-Digitized Sky 
Survey and its size is 40\arcmin$\times$ 40 \arcmin. North is at the top, 
East to the left.  }
\end{figure*}  
 
\subsection{Data Reduction} 
The data were reduced using the IRAF multi-fibre spectra reduction package 
DOFIBER.  
The sky subtraction represented the most difficult part of the
reduction.  As already mentioned, we used some of the fibres to
monitor the sky background during the science exposures and we also
had several offset sky exposures with the same fibre configuration.
We used the sky fibres in our exposures of M~33 to monitor the
relative intensity of the atmospheric emission lines, which are known
to vary during the night.  However, these sky fibres lie in different
positions in the spectrograph focal plane and consequently produce
spectra with different spatial and spectral instrumental profiles.
For this reason, they cannot be used for a direct sky subtraction.
There are instead no variations in the instrumental profiles for each
individual fibre between the offset sky frame spectra and the science
frames, as fibres are fixed at the spectrograph entrance.  We
therefore used these offset sky spectra to do the sky subtraction,
after correcting the relative intensity of the atmospheric lines using
the mean sky spectrum computed from the sky fibres observed at the
same time as the science targets.
 
\subsection{Flux calibration} 

Relative data flux calibration was obtained taking spectra with
several fibres of the spectrophotometric standard stars G191B2B (Oke
\cite{oke}) and 40 Eri B (Oke \cite{oke74}).  Using AF2, the absolute
flux calibration cannot be done, since one is limited by the fixed
fibre diameter.  In addition, it is impracticable to observe a
spectrophotometric standard star in all the science fibres in order to
have a sensitivity function for each fibre.  In order to test whether
a mean sensitivity function can be used for all the fibres, the
standard star G191B2B was observed in nine fibres and the sensitivity
functions of those fibres were analyzed.  The variations of the
spectral response of each fibre are within $\sim$1\% in the whole
spectral range, as shown in Figure 2.  
This result allows us to use a mean sensitivity function for all the
fibres.
 
\begin{figure}  
\caption{The percentage variation from fibre to fibre of the  sensitivity  
function.  }  
\label{fibers} 
\resizebox{\hsize}{!}{\includegraphics{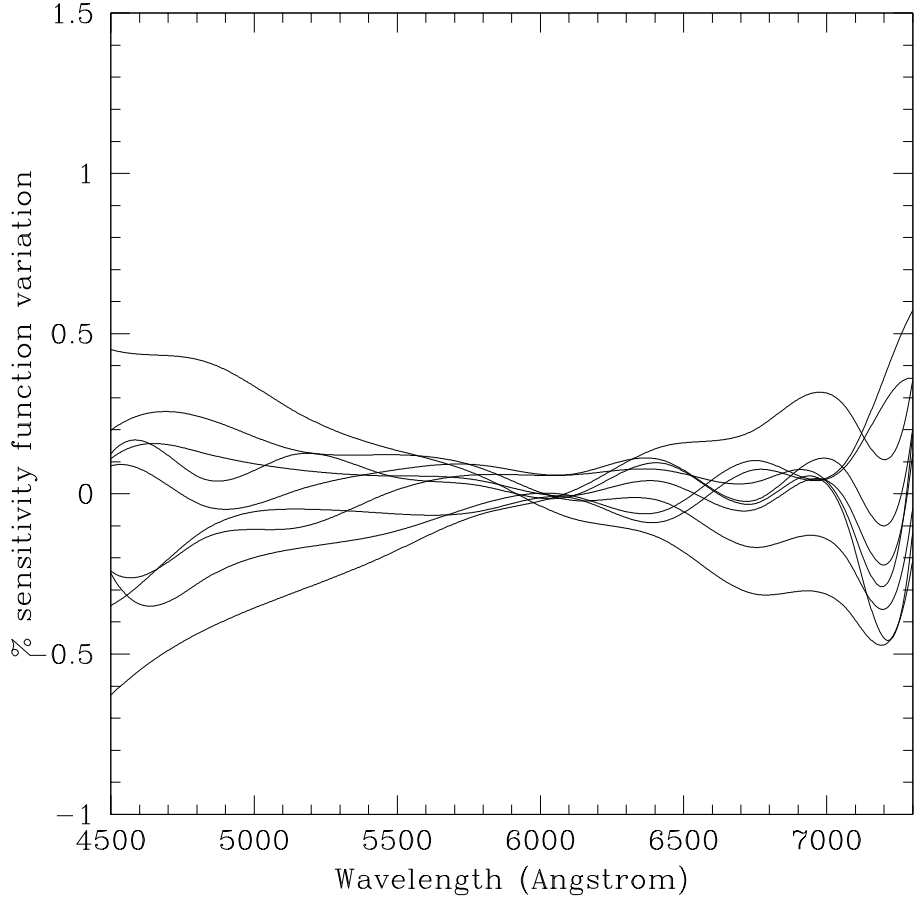}}  
\end{figure}  
  
\subsection{Flux measurement and de-reddening} 

Emission line fluxes of the spectra were measured using the
MIDAS/ALICE package.  We also used the IRAF task TWOFITLINES
(IACTASKS, Acosta \cite{acosta}) to estimate errors on the fluxes.
This task computes the uncertainty of model parameters by a number of
repeated fits to simulated data, generated from the best model with
random noise added.  The absolute errors on the measurement of the emission-line 
strengths are given; percentage errors (not including flux calibration errors) on the
\hb\ line are shown in Table \ref{flux1}.  
The background noise and the sky subtraction are the greatest sources
of error and they affect uniformly the whole spectral range where
emission lines were measured.  Thus, the absolute error of each
emission line of our spectra is approximately constant and equal to
the absolute error of \hb\ line.
Moreover, systematic errors due to the flux calibration, 
including the errors on the Oke fluxes ($\sim$ 0.01 mag), were  estimated to be 
approximately 5-10\% along the whole spectral range.
 
The observed line fluxes have been corrected for the effect of the
interstellar extinction.  The extinction law of Mathis (\cite{mathis})
with $R_V$=3.1 has been used.  The \cbeta, which is the logarithmic
difference between the observed and un-reddened \hb\ fluxes, was
determined comparing the observed Balmer I(\ha)/I(\hb) ratio with
 its theoretical value.  All fluxes are presented on a scale relative
to \hb, where F(\hb) is set to 100.  Observed fluxes and \cbeta\ for
the objects in whose spectra we could extract information are shown
in Table \ref{flux1}.  The percentage errors on the \hb\ fluxes are
presented in the last line of Table \ref{flux1}.
 
\begin{table*} 
\caption{Observed line fluxes. }   
\label{flux1} 
\begin{center} 
{\scriptsize
\begin{tabular}{l l r r r r r r r}     
\hline    
Identification & &    MCMP8 & MCMP9  & MCMP17 & MCMP18 & MCMP23 & MCMP25&   
MCMP28\\
	      & &    PN         & PN           & -               & PN            & \hii\ r     & PN           & 
PN\\	
\hline  
4685.7 & \heii 	&-&-&-&-&-&-&29.1\\ 
4861.3 & \hb 	&100.&100.&100.&100.&100.&100.&100.\\ 
4958.9 & \oiii 	&160.&467.&206.&255.& 67.3& 221.&396. \\ 
5006.8 & \oiii 	&454.&1275.&469.&641.& 165.& 803.& 1198. \\ 
5875.7 & \hei 	&-&33.9&-&-&22.0&- &13.6\\ 
6562.8&\ha	&636.&741.&285.&350.&364.&446.&509.  \\ 
6583.4&\nii	&52.3&108.&-&75.3&29.8&61.0&45.3  \\ 
6678.1&\hei	&14.8&20.8&-&20.3&-&-&11.5\\ 
6716.5 &\sii	&23.9&17.5&18.3&65.0&70.6&132.&50.2 \\ 
6730.8&\sii	&23.9&21.6&18.8&39.2&50.&-&36.6\\ 
7065.3&\hei	&23.1&-&-&35.0&-&-&-\\ 
7135.8 &\ariii	&62.7&80.2&-&62.8&-&-&48.9\\ 
7325 (*) & \oii	&22.7&15.2&-&-&-&-&19.0 \\ 
\cbeta & 		&1.15&1.36&0.00&0.30&0.35&0.64&0.83\\ 
\hb\ error   &		&2\%&10\%&15\%&4\%&8\%&50\%&6\%\\ 
\hline    
 & &  MCMP31 & MCMP38  & MCMP40& MCMP41 & MCMP42 & MCMP45 &  MCMP49\\  
& & \hii\ r/PN & PN	        & PN	         &  SNR         & SNR          & PN            & PN\\
\hline    
4685.7 & \heii 	&13.6&-&-&30.6&-&-&16.5\\ 
4861.3 & \hb 	&100.&100.&100.&100.&100.&100.&100.\\ 
4958.9 & \oiii 	&197.&134.&517.&119.&339.&218.&208. \\ 
5006.8 & \oiii 	&314.&158.&980.&265.&840.&423.&487. \\ 
5875.7 & \hei 	&15.4&91.9&-&64.6&-&14.6&21.4 \\ 
6562.8&\ha	&288.&285.&662.&736.&329.&296.&595.\\ 
6583.4&\nii	&103.&71.9&964.&250.&130.&96.1&88.8\\ 
6678.1&\hei	&13.3&-&-&-&-&-\\ 
6716.5 &\sii	&41.0&28.4&183.&189.&284.&26.6&53.0\\ 
6730.8&\sii	&30.7&-&135.&158.&123.&23.6&32.7\\ 
7135.8 &\ariii	&32.6&-&-&75.1&-&15.9&-\\ 
\cbeta 		&&0.01&0.00&1.20&1.36&0.20&0.05&1.05\\ 
\hb\ error   &		&8\%&15\%&40\%&20\%&50\%&10\%&8\%\\ 
\hline 
 & &  MCMP53 & MCMP60  & MCMP61 & MCMP65 & MCMP68 & MCMP69&  MCMP71\\  
&&    SNR         & PN             & \hii\ r/PN & PN           & PN            & PN            & PN\\
\hline    
4363.2 & \oiii 	&-&16.4&-&20.2&-&- 		&-      \\ 
4685.7 & \heii 	&-&-&78.2&-&-&-		&-      \\ 
4740.2& \ariv	&-&-&-&-&-&13.4			&-      \\ 
4861.3 & \hb 	&100.&100.&100.&100.&100.&100.	&100.\\ 
4958.9 & \oiii 	&-&568.&274.&510.&160.&288. 	&365.\\ 
5006.8 & \oiii 	&-&1475.&637.&1259.&520.&748. 	&553.\\ 
5875.7 & \hei 	&-&-&-&-&32.4&41.9 		&-\\ 
6562.8&\ha	&1177.&484.&320.&567.&365.&555.	&658.\\ 
6583.4&\nii	&428.&221.&98.1&163.&134.&693.	&539.\\ 
6716.5 &\sii	&410.&54.1&43.3&71.7&67.2&201.	&53.7\\ 
6730.8&\sii	&356.&48.6&49.7&74.8&63.3&163.	&63.0\\ 
7065.3&\hei	&-&-&-&22.5&16.2&68.2		&-      \\ 
7135.8 &\ariii	&-&52.0&84.8&119.&39.1&145.	&-      	\\ 
7325(*) & \oii	&-&29.6&-&40.&-&-		&-\\ 
\cbeta & 		&2.03&0.76&0.17&0.97&0.35&0.95	&1.19\\ 
\hb\ error   &		&50\%&20\%&20\%&10\%&8\%&20\%&50\%\\ 
\hline 
\end{tabular}    
}
\end{center} 
Note: (*) Sum of the \oii\ doublet  lines at  7319.6 and  7330.2 \AA. 
\end{table*}

\begin{table*}  
{\bf Table 1} {\sl - continued} 
\begin{center} 
{\scriptsize
\begin{tabular}{l l r r r r r r r }     
\hline 
Identification & &  MCMP75 & MCMP77  & MCMP78 & MCMP91 & MCMP93 & MCMP96&   MCMP101\\ 
&&                             PN           & -                & \hii\ r      & PN            & PN           & 
PN           & PN \\
\hline    
4363.2 & \oiii 		&-&-&-&9.6&6.09&-&10.0 \\ 
4471.5 & \heii 		&-&-&7.53&12.3&14.7&-&- \\ 
4685.7 & \heii 		&-&-&-&18.8&8.52&-&-\\ 
4740.2& \ariv		&11.9&-&-&-&-&-&-\\ 
4861.3 & \hb 		&100.&100.&100.&100.&100.&100.&100.\\ 
4958.9 & \oiii 		&451.&265.&51.1&448.&406.&404.&460. \\ 
5006.8 & \oiii 		&1212.&765.&157.&1348.&1200.& 1199.&1385. \\ 
5875.7 & \hei 		&-&-&23.9&-&-&19.9&24.3 \\ 
6562.8&\ha		&366.&600.&427.&433.&430.&438.&499.\\ 
6583.4&\nii		&151.&-&73.2&153.&91.8&96.6&74.1\\ 
6678.1&\hei		&17.3&-&8.66&11.0&5.95&10.1&11.4\\ 
6716.5 &\sii		&10.4&-&50.1&9.4&19.5&33.0&15.2\\ 
6730.8&\sii		&18.7&-&35.2&13.2&21.6&22.5&11.3\\ 
7065.3& \hei		&24.6&-&7.32&10.5&-&23.&-\\ 
7135.8 &\ariii		&45.3&-&23.7&31.2&27.3&36.5&38.6\\ 
7319.6 & \oii		&-&-&12.1&10.5&10.8&-&12.3\\ 
7330.2 & \oii		&-&-&10.9&7.7&8.3&-&10.8\\ 
\cbeta & 			&0.36&1.06&0.58&0.60&0.59&0.62&0.80\\ 
\hb\ error   &			&8\%&50\%&9\%&2\%&5\%&5\%&3\%\\ 
\hline 
 & &  MCMP105 & MCMP106  & MCMP122 & MCMP125 & MCMP127 & MCMP130& MCMP132\\ 
&& PN                 & PN                & \hii\ r/PN  & PN               & \hii\ r       & -                & 
\hii\ r/PN \\
\hline    
4363.2 & \oiii 	&-&-&1.36&-&-&-&-			\\ 
4471.5 & \heii 	&-&-&8.68&-&-&20.5&7.93			\\ 
4861.3 & \hb 	&100.&100.&100.&100.&100.&100.&100.		\\ 
4958.9 & \oiii 	&367.&228& 58.0&189.&61.8&50.9&90.8		 \\ 
5006.8 & \oiii 	&787.&265.&179.&583.&194.&124.&234.		 \\ 
5875.7 & \hei 	&-&-&25.4&14.9&27.0&18.0&23.5		 \\ 
6562.8&\ha	&382.&785.&689.&458.&755.&395.&621.		\\ 
6583.4&\nii	&43.6&476.&151.&123.&221.&160.&88.7		\\ 
6678.1&\hei	&-&-&10.9&-&77.3&-&21.0			\\ 
6716.5 &\sii	&18.2&81.3.&61.6&53.8&148.&-&40.8		\\ 
6730.8&\sii	&-&55.1&55.8&35.9&119.&-&36.9		\\ 
7065.3&\hei	&-&-&10.5&-&-&-&21.7			\\	 
7135.8 &\ariii	&15.3&-&29.7&34.1&95.5&-&50.4		\\ 
7319.6 & \oii	&-&-&10.0&-&-&-&-			\\ 
7330.2 & \oii	&-&-&10.3&26.9*&-&-&-			\\ 
\cbeta & 	&0.42&1.45&1.26&0.68&1.39&0.47&1.11		\\ 
\hb\ error   &		&12\%&40\%&2\%&5\%&12\%&10\%&4\%\\ 
\hline 
 & &MCMP134& MCMPc5 & MCMPc6  & MCMPc18 &  MCMPc35 & MCMPc43&MCMPc50 \\  
&  & PN	      & \hii\ r      & \hii\ r      & SNR            & \hii\ r         & SNR          & -\\
\hline    
4685.7 & \heii 	&28.5&-&-&-&-&-&-\\ 
4861.3 & \hb	&100.&100.&100.&100.&100.&100.&100.\\ 
4958.9 & \oiii 	&270.&-&27.4&-&21.5&23.0&- \\ 
5006.8 & \oiii 	&703.&33.6&101.&-&72.0&30.3&9.73 \\ 
5875.7 & \hei 	&-&-&7.14&-&9.90&-&- \\ 
6562.8&\ha	&490.&632.&473.&371.&476.&394.&738.\\ 
6583.4&\nii	&173.&78.6&49.4&55.8&97.8&174.&-\\ 
6678.1&\hei	&32.2&-&3.90&-&5.62&-&-\\ 
6716.5 &\sii	&60.6&96.7&42.7&136.&41.2&147.&-\\ 
6730.8&\sii	&51.2&56.4&29.2&99.0&29.0&110.&-\\ 
7065.3&\hei	&-&-&5.13&-&4.93&-&18.8\\	 
7135.8 &\ariii	&64.7&-&7.94&-&10.1&-&-\\ 
\cbeta & 	&0.78&1.14&0.73&0.38&0.73&0.46&1.36\\ 
\hb\ error   &		&20\%&10\%&1\%&25\%&0.3\%&7\%&3\%\\ 
\hline 
\end{tabular}
}    
\end{center} 
Note: (*) Sum of the \oii\ doublet  lines at  7319.6 and  7330.2 \AA. 
\end{table*}    
  
\section{Analysis of the spectra} 
 
Out of the 48 targets observed, the spectra of 3 candidate PNe and 2
unclassified objects were so weak that no usable information could be
extracted, whereas one of the unclassified objects, namely MCMPc9,
turned out  to be a normal star.  The nature and properties
of the remaining 42 objects are discussed in the following.

\subsection{The Balmer decrement: PNe extinction} 
 
The weighted mean value of the logarithmic extinction, \cbeta, of the
PN candidates (Table 1) is $0.73\pm 0.21$. Approximately 30\% of the
candidates have a very small \cbeta\ and presumably they lie in  front
of the galaxy.  The other PNe have
\cbeta\ greater than 0.5: they lie on the far side of M~33 and/or in
regions of the spiral arms where the extinction is higher.  The mean
logarithmic extinction of our disc PNe is similar to that found by
JC99 in three PNe on the disc of M~31 (\cbeta = 0.50; JC99).  
 
\subsection {Diagnostic diagrams} 
\label{diag} 
\begin{figure}  
\label{canto} 
\caption{Diagnostic diagram (originally from Sabbadin \& D'Odorico,  
\cite{sd76}).   
The Galactic \hii\ regions and SNRs areas are from Garc\'{\i}a-Lario et al.  
\cite{garcia}.  
The PNe region is drawn using recent data of Galactic and
extragalactic PNe (see text). Our candidate PNe are represented by
triangles (R$>$1.6), circles (0.3$<$R$<$1.6) and squares
(R$<$0.3). The emission-line objects with non-negligible continuum are
represented by stars. }
\resizebox{\hsize}{!}{\includegraphics{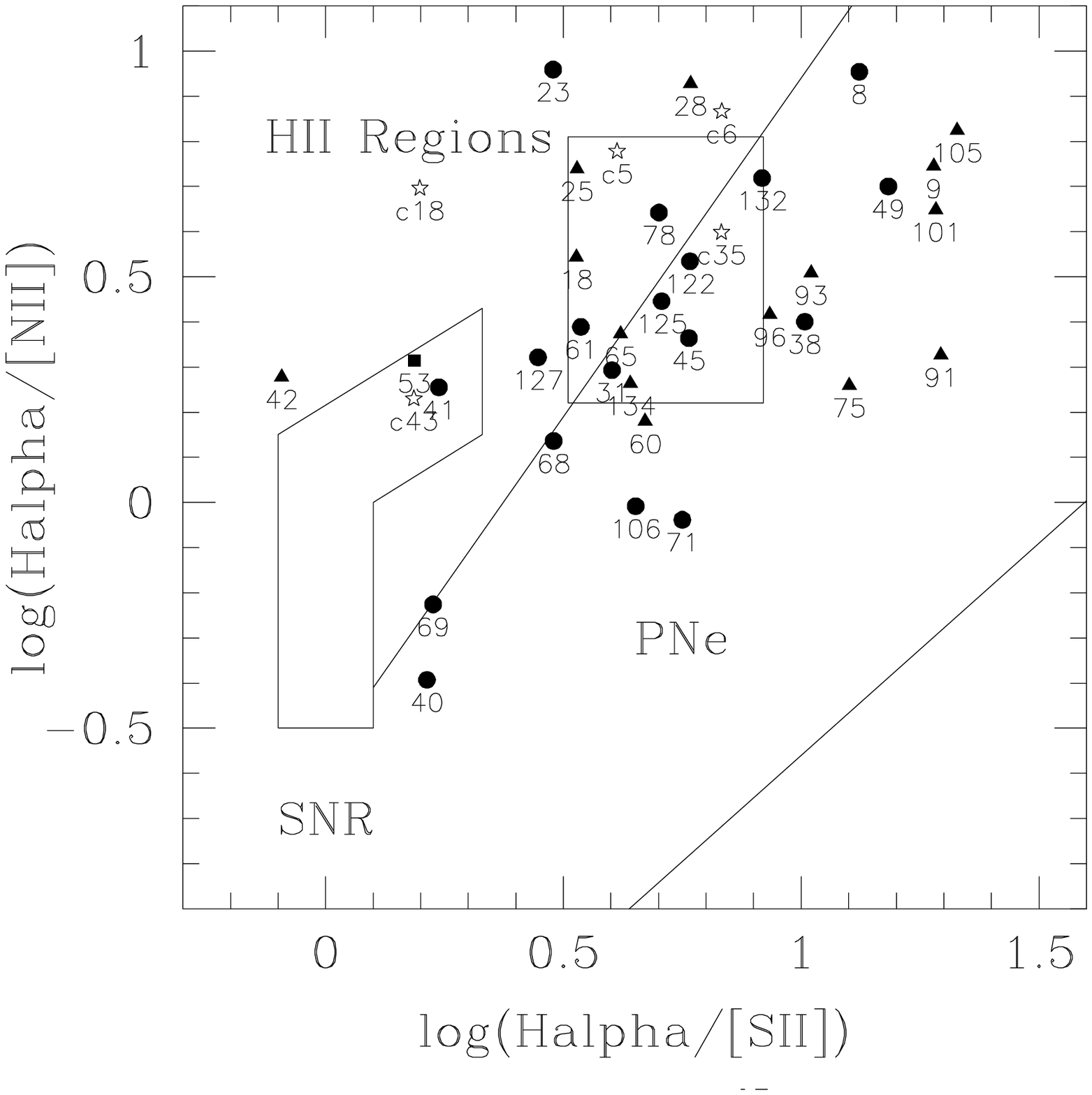}}  
\end{figure}  
 
Magrini et al. (\cite{magrini00}) introduced the excitation parameter
R=${\frac{\rm [OIII]}{H\alpha + \rm[NII]}}$ as a crude indicator of
the nature of extragalactic emission line objects found in narrow-band
surveys, arguing that most objects with $R>1$ (and point-like at the
distance of M31 and M33) are expected to be PNe.  Ciardullo et
al. (\cite{ciardullo02}) suggest a slightly higher limit ($R>1.6$) for
PN candidates in the brightest 1-magnitude bin of the PNLF.  
Nevertheless, using $R>1.6$, we would exclude a large number of low excitation 
PNe. In fact about  40\% of confirmed Galactic 
PNe have  $R<1.6$ (see M00, Figure 3). 
In the following analysis on the nature of candidate PNe, we consider  adequate for  PNe 
values of $R$ from 0.3 to larger than 1.6, as expected for a `normal' (i.e. Galactic) population of PNe. 
However, to identify {\sl bona fide} PNe, more precise indicators should be used when, as in our case, 
a larger part of the  spectrum is available.
The empirical emission line diagrams (Sabbadin \& D'Odorico
\cite{sd76}, Garc\'{\i}a-Lario et al. \cite{garcia}, Riesgo-Tirado \&
L\'opez \cite{rt02}) have proved to be very useful in distinguishing among
different classes of Galactic ionized nebulae, the most effective one
being probably that displaying the \ha/\nii\, vs. \ha/\sii\ line
ratios, as it neatly separates PNe from \hii\ regions and supernova
remnants.  This diagnostic diagram is especially valuable for
extragalactic objects since it involves relatively strong lines and
avoids information less easily obtainable in the UV or IR spectral
ranges.  We have recalculated the PNe zone in this diagram using the
most precise Galactic and extragalactic data available: Kingsburgh \&
Barlow (\cite{kb1994}, hereafter KB94), Kaler et al. (\cite{kaler}),
Cuisinier et al. (\cite{cuisinier}) for Galactic PNe; de Freitas Pacheco
et al. (\cite{defreitas}), Vassiliadis et al. (\cite{vass92}) 
and Stanghellini et al. (\cite{stanghe}) for LMC and SMC; 
JC99 and Stasinska et al. (\cite{stas1998}) for M~31; Walsh 
et al. (\cite{walsh97}) for Fornax.  Note 
that the new limits shift the upper border of the PNe band more into
the \hii\ region box.
 
The M~33 PN candidates were put in this diagram (Figure 3) coded by
their $R$ parameter.  In this way we supplement the diagram with the
information of the \oiii\ line which is the most important optical
line characterizing a PN.  We analyze in the following the location of
all objects in this diagram.
Five of the nine objects with non-negligible continuum (see M00) have
good data for the involved lines (Table 1).  Two of them (MCMPc18 and
43) are near the SNRs locus; both have strong \sii\ lines (in MCMPc18
the \sii\ lines are stronger than \nii). The remaining three objects
appear in the \hii\ region area. All of them are associated with a low
value of the excitation indicator ($R<0.5$).  Only one is located in
the overlapping area between the PNe and the \hii\ regions (MCMPc35)
but it is indeed a low excitation object and thus presumably it is not
a planetary nebula.
Most objects without continuum were found to occupy the PNe region
although they appear slightly displaced towards the upper border of the
PNe band.  Note, on the other hand, that the {\sl three}  objects where
chemical abundances were determined (Section 4) lie inside the PNe
area, and that all candidates in the PNe area show values of $R$ from
0.3 to larger than 1.6, as expected for a normal
population of PNe.  In fact, only three candidates in the overlapping
PNe and \hii\ region area have $0.5<R<1$ (MCMP31, MCMP122, MCMP132).
Although those objects can be misidentified with
\hii\ regions, they can equally well be low excitation PNe. The
remaining candidates can be considered {\sl bona fide} planetary
nebulae.
A total of ten candidates lie above the line plotted to mark the PNe
region: seven in the \hii\ region area and three in the SNR
area. Three of the former, MCMP23, MCMP78 and MCMP127, have
$R<0.5$ and are probably unresolved \hii\ regions.  MCMP61 has higher
excitation and its nature remains doubtful.  The other three objects
(MCMP18, MCMP25, MCMP28) have a high $R$ and lie near the PNe locus
where there are also numerous confirmed PNe.  They are probably
genuine PNe.  MCMP41, MCMP42 and MCMP53 occupy the SNRs locus, each
with different excitation.  

Concluding, no emission line object with some continuum emission
(called  'unclassified ' in our previous work; M00) can be considered
to be PNe, whereas most of our candidate PNe (objects without
continuum) are shown to occupy the same locus in the \ha/\nii\,
vs. \ha/\sii\ diagnostic diagram as Galactic and extragalactic PNe.
Three candidates which lie in the upper \hii\ region area and have a
low value of $R$ are rejected as PNe, whereas four objects in that
area showing high $R$ are classified as PNe.  Three candidates located
in the SNRs region are rejected as PNe and another three candidates in
the overlapping area between PNe and \hii\ regions can be low
excitation PNe or compact \hii\ regions. All in all, 72\% of our
candidate PNe with good spectral data (i.e. with \ha, \nii,
\sii. \oiii\ lines well measured) are {\sl bona fide} PNe; the
remaining 28\% will include \hii\ regions, SNR and low excitation PNe.
From the galaxy distribution of $R$ discussed in our previous work
(M00), we did expect a contamination by \hii\ regions of approximately
11\%, a figure which is in rough agreement with the values found
above.  The probable nature of each object, as deduced from the 
preceding discussion, is specified in Table 1.
   
\subsection{Electron density and temperature} 
 
The essential line for electron temperature determination, 4363\AA\
\oiii, was measured in 6 PNe ( MCMP60, 65, 91, 93, 101, 122) .  
For the other 36 objects, it was not 
observed: for 21 objects it was too faint to be detected while in the 
remaining 15 the spectral range did not include it (Section 2). 
The spectra of PNe MCMP91, 93, 101 and 122 had good signal to 
noise ratio, whereas MCMP60 and 65 were quite noisy. 
To test {\bf the} quality of the spectra of the PNe where physical and chemical quantities can 
be computed, 
we compared the observed and theoretical values for the line ratios set 
by atomic constants and present in our spectral range: \oiii\ I(5007)/I(4959) and \nii\ I(6584)/I(6548).
The observed mean for the \oiii\ ratio is 2.85$\pm$ 0.40 and for 
the \nii\ ratio is 2.85$\pm$ 0.50, 
considering the sample of  six nebulae, 
and 3.01$\pm$0.10 and  3.02$\pm$0.30 for  MCMP91, 93, 101 and 122, 
whereas the theoretical value, corrected for the energy of the levels, 
are respectively 2.96 and 2.98 (Osterbrock \cite{ost1989}). 
We computed electron temperature and density for the whole six nebulae sample, 
avoiding  chemical abundance determinations 
for the two PNe whose spectra were noisier,
 as deduced from their observed 
\oiii\ and \nii\ line ratios and from their overall low S/N and 
taking out also the object MCMP122 whose 
classification as a PN was doubtful.   

Electron temperatures and densities have been derived using the ratios
\oiii\  I(5007)/I(4363) for T$_e$ and \sii\ I(6717)/I(6731) for n$_e$.  
The most relevant source of error for the temperature-sensitive ratio
is the error associated with the weak \oiii\ 4363\AA\, line, whereas
the error on the density ratio is based on the accuracy of the fluxes
of both lines of the \sii\ doublet.  
The uncertainty on T$_e$ is approximately 15\% for 
the good S/N spectra, and somewhat higher for the remaining ones.  
Errors in the electron density are larger but  density 
essentially does not enter into the chemical abundance determinations.
We adopted T$_e$\nii=T$_e$\oiii,
i.e.  the same temperature for the high and low ionization ions,
because the \nii\ 5755\AA\ line was not detected in our spectra and
consequently T$_e$\nii\ could not be determined directly from the
measurement of the \nii\ I(5755)/I(6583) ratio.  The nebular electron
temperatures and densities for the 6 PNe where we could measure the
\oiii\ 4363\AA\ line are presented in Table
\ref{tempdens}.  In this Table their distances in kpc from the centre of 
M~33, their mag$_{\rm [OIII]}$ (from M00, extinction corrected with
the \cbeta s of Table \ref{flux1}), and their de-reddened excitation
parameter R=${\frac{\rm [OIII]}{H\alpha + \rm[NII]}}$ are also
indicated.
 
\begin{table}    
\caption{Electron densities and temperatures. }   
\label{tempdens} 
\begin{center} 
\begin{tabular}{ l l l l l }     
\hline    
 	& {\small MCMP60} &{\small MCMP 65} &{\small MCMP 91}\\
\hline 
n$_e$ (cm$^{-3}$)		&300	&700	&1900	 \\ 
T$_{[\rm{O III}]}$  (K) 	&12600	&14900	&11100	\\ 
d (kpc) 			&3.1	&1.8	&4.3	\\ 
mag$_{\rm [OIII]}$ 		&22.50	&21.98	&20.78 \\ 
R			&3.2	&3.2	&3.2	\\ 
&&&&\\ 
\hline 
&  			 {\small  MCMP93}  	& {\small  MCMP101} & {\small  MCMP122} &   \\  
\hline 
n$_e$  (cm$^{-3}$)		&850	&100	&400	\\ 
T$_{[\rm {OIII}]}$  (K) 	&9900	&11500	&13200	&	\\ 
d (kpc)			&3.8	&3.4	&0.8	 \\ 
mag$_{\rm [OIII]}$		&21.64	&21.19	&21.30	\\ 
R			&3.2	&4.1	&0.5	\\ 
&&&&\\ 
\hline 
 
\end{tabular}    
\end{center} 
\end{table}    
 
\subsection{Chemical abundances} 
 
As mentioned before, chemical abundances could be derived with a
reasonable accuracy only for  three out of the six PNe, 
namely MCPM91, 93, and 101, in whose spectra the \oiii\
$\lambda$4363\AA\ was detected  since their S/N was good also for the weakest lines.

Helium ionic abundance have been 
computed considering the Case B recombination and using the effective
recombination coefficients from Hummer \& Storey (\cite{hs1987}) for
\hb\ and
\he$^{2+}$ and from Brocklehurst (\cite{bro71}) for \he$^{+}$.  The
\hei\ line strengths were corrected for the effect of collisional
population of their upper state following Clegg (\cite{clegg}). The
contribution of this effect is negligible in the case of low densities.
The ionic metal abundances were derived from their collisionally
excited lines solving the equations of statistical equilibrium.  In
order to calculate the total abundances, the unseen stages of
ionization were accounted for using the standard ionization correction
factors (ICF) (KB94).  We used the same atomic data as KB94.\\
We estimated errors in the chemical abundances of these three objects on 
the bases of the error propagation across the procedure of evaluation of abundances, 
taking into account the uncertainties in the observed fluxes of the 
relevant lines and in the electron temperatures. 
Errors due to the uncertainties in the atomic quantities were not  considered 
due to the extreme complexity of that, also regarding obvious compensation 
of these errors among the various atomic quantities. 
We  obtained formal errors of: 20-30\% for He, 25-50\% for O, 40-70\% for N, and 30-70\% for Ar. 
Errors are not given for S  since its abundance is uncertain 
and it will be discussed in Section 4.2. 
The derived chemical abundances of the three PNe are
listed in Table~\ref{chemim33}.  
 
\begin{table}    
\caption{Chemical abundances of three  PNe in M~33. }   
\label{chemim33} 
\begin{center} 
\begin{tabular}{l  l l l l }     
\hline 
    
\hline 
Ion/Elem.	  &  MCMP91   & MCMP93  & MCMP101    \\  
\hline 
He$^{+}$/H  				&0.17	&0.09	&0.14			\\ 
He$^{2+}$/H  				&0.02	&0.01	&-			\\ 
{\bf He/H}	 			&0.19	&0.10	&0.14	    		\\ 
O$^{+}$/H$\times$10$^4$  		&1.44	&2.37	&2.96			\\ 
O$^{2+}$/H$\times$10$^4$ 		&3.25	&4.20	&3.03			\\ 
{\sl icf} 				&1.08	&1.05	&1.00			\\ 
{\bf O/H}$\times$10$^4$  		&5.06	&6.92	&5.72			\\ 
N$^{+}$/H$\times$10$^5$   		&1.90	&1.22	&0.77			\\ 
{\sl icf} 				&3.51	&2.92	&2.12			\\ 
{\bf N/H}$\times$10$^4$  		&0.67	&0.30	&0.16			\\ 
Ar$^{2+}$/H$\times$10$^6$ 		&1.49	&1.51	&1.48			\\ 
{\sl icf} 				&1.40	&1.52	&1.89			\\ 
{\bf Ar/H}$\times$10$^6$  		&2.08	&2.29	&2.80			\\ 
S$^{+}$/H$\times$10$^6$ 		&0.49	&0.85	&0.36			\\ 
{\sl icf} 				&1.16	&1.12	&1.06			\\  
{\bf S/H}$\times$10$^6$ 			&0.57	&0.92	&0.38			\\ 
\hline 
\end{tabular}    
\end{center} 
\end{table}

\section{Discussion on the chemical abundances} 
 
\subsection{Comparison with other galaxies} 
 
\begin{table}    
\caption{ M~33 PNe chemical abundances in comparison with other abundances,  
with (N, O, S, Ar) expressed in the usual $\lg[\frac{X}{H}] + 12$ units. } 
\label{median} 
\begin{center} 
\begin{tabular}{l  l  l l l l l}     
\hline 
 &Galaxies	&			He/H&	N/H&	O/H&	S/H&	Ar/H\\ 
\hline	 
	&M~33$^{(a)}$ &		0.14&	7.50&	8.76&	5.76&	6.36\\ 
	&M~31$^{(b)}$&		0.13&	8.03&	8.35&	7.38&	6.00\\ 
	&Galaxy$^{(c)}$&		0.115&	8.35&	8.68&	6.92&	6.39\\ 
PNe	&LMC$^{(d)}$&		0.105&	8.07&	8.44&	6.51&	5.93\\ 
	&SMC$^{(d)}$&		0.107&	7.84&	8.24&	6.48&	5.62\\ 
	&Sgr$^{(e)}$&		0.106&	7.41&	8.30&	6.40&	5.82\\ 
	&Cen~A$^{(f)}$&		-&	8.02&	8.39&	7.10&	-\\	 
\hline	 
	&M~33$^{(g)}$&		0.097&	6.98&	8.13&	6.81&	5.85\\ 
	&M~33$^{(h)}$&		0.083&	7.34&	8.55&	6.95&	-\\ 
	&M~31$^{(i)}$&		-&	7.63&	8.86&	7.11&	-\\ 
\hii\  	&Orion$^{(j)}$&		0.100&	7.83&	8.60&	6.93&	6.65\\ 
regs.	&Galaxy$^{(k)}$&		0.100&	7.57&	8.70&	7.06&	6.42\\	 
	&LMC$^{(k)}$&		0.085&	6.97&	8.43&	6.85&	6.20\\	 
	&SMC$^{(k)}$&		0.080&	6.46&	8.02&	6.49&	5.78\\ 
\hline 
	&Solar$^{(d)}$&		0.098&	8.00&	8.93&	7.21&	6.56\\ 
 
\hline 
\end{tabular}    
\end{center} 
$^{(a)}$This work (median values of 3 PNe) 
$^{(b)}$Jacoby \& Ciardullo (\cite{jacoby1999}),  
$^{(c)}$Kingsburgh \& Barlow (\cite{kb1994}),  
$^{(d)}$Clegg (\cite{clegg92}),  
$^{(e)}$Walsh et al. (\cite{walsh97}),  
$^{(f)}$Walsh et al. (\cite{walsh99}),  
$^{(g)}$Kwitter et al. (\cite{kwitter}),  
$^{(h)}$V\'{\i}lchez et al (\cite{vilchez88}) ,  
$^{(i)}$Blair et al. (\cite{blair}),  
$^{(j)}$Rubin et al. (\cite{rubin}),  
$^{(k)}$Dufour (\cite{dufour}). 
 
\end{table}    
 
The median abundances of Helium, Nitrogen, oxygen, sulphur and argon
of the  three  PNe are given in Table \ref{median}.  He abundance is given
as He/H, whereas (N, O, S, Ar) abundances are in the usual
$\log[\frac{X}{H}] + 12$ units.  
In Table \ref{median} we show for comparison chemical abundances of 15 PNe
in M~31 (JC99), the Milky Way (from the sample of KB94), the LMC and
the SMC (Clegg \cite{clegg92}), the Sagittarius dwarf galaxy (Walsh et
al. \cite{walsh97}) and the elliptical galaxy Centaurus A (NGC~5128,
Walsh et al. \cite{walsh99}). These data include all the most recent
information concerning chemical abundances of extragalactic PNe (Ford
et al \cite{ford}).  We also present the abundances of several \hii\
regions: of M~33 (Kwitter \& Aller \cite{kwitter}, V\'{\i}lchez et
al. \cite{vilchez88}), of M~31 (Blair et al. \cite{blair}), of LMC and
SMC (Dufour \cite{dufour}) and of the Orion nebula (Rubin et
al. \cite{rubin}).  Finally, the figures can be compared with the
solar abundances (Clegg \cite{clegg92}).
 
The median helium abundance of our PNe is  close to  that
of the PNe in M~31 (JC99), and slightly higher than the Milky Way
PNe one.  The oxygen and argon abundances are in agreement with
the values obtained by KB94.  They are typically lower than the solar
values, as observed in Galactic PNe.  The nitrogen and sulphur
abundances in M33 PNe are unexpectedly low.
 
A first comment comes from the comparison of PNe and \hii\ region chemical abundances (see Table \ref{median}).  The stellar evolution
theory predicts that PNe should have enhanced abundances of He, N and
C whereas the elements heavier than N should have abundances not
appreciably different from those at the moment of the formation of
their progenitor stars (cf. Iben \& Renzini 1983).

The comparison of abundances in PNe and \hii\ regions for the galaxies
of Table 4 shows that in our Galaxy, LMC and SMC the theoretical
predictions in all elements are verified. The
prediction appears to be verified for He and N in M33 and for N in M31 
(note that in M31 the \hii\ region abundance of He is not reported) 
whereas for O, S, and Ar the numbers do not support the predictions.  
Since the abundances both of
PNe and \hii\ regions are better measured in our Galaxy and the MCs
than in M31 and M33, we infer from the above that the abundances of O,
S, Ar in M31 and M33 might not be accurate enough to claim agreement or
disagreement with the theory as far as the behaviour in PNe versus
\hii\ regions is concerned.  
 We also compare the N/O of the three PNe in M33 to that found in \hii\ regions in the same galaxy.
According to Table \ref{median}, N/O is, on average, 0.05 for the three PNe, and
similar to the N/O ratios of 0.07 and 0.06 found by Kwitter et al. \cite{kwitter} and
V\'{\i}lchez et al. \cite{vilchez88} for \hii\ regions in M33. 
For any other galaxy in Table \ref{median}, N/O for PNe is
much larger than for \hii\ regions, showing the N enrichment occurring in the
atmospheres of intermediate mass stars. 
It seems that such an enrichment has not occurred for the three PNe in M33. 

In the following, we first consider the abundance of S, which is 
unexpectedly low both in comparison with that in the \hii\ regions of
M33 and relative to PNe in our Galaxy.  Next we address  the abundance
of N, which is higher than the corresponding abundance in the
\hii\ regions of M33, but lower than PNe in the Galaxy. 

\subsection {The sulphur  abundance} 
 
The unexpected low sulphur abundance in the PNe observed in M~33 can be
explained by the fact that we are not observing the \siii\ lines. In
high excitation objects, such as PNe, most of the sulphur is twice ionized,
and abundances derived using only \sii\ lines are not representative  of
its total abundance, even if corrected with the nominal ICF.  
 Thus we must consider our value as a lower limit to the true abundance 
of S.
 

   
\subsection {The nitrogen abundance} 
 
The nitrogen abundance that we found is quite low for a galaxy of
normal metallicity, as M~33 is, if the abundances of the (best observed)
elements like He and O are compared with the corresponding values for
the Galactic PNe. Our N/H median is 7.5, which is 0.85 dex below the
Galactic value, while the O/H abundances are comparable (see Table
\ref{median}).  We have investigated various possible reasons of this 
``under-abundance''.  First, our chemical abundance determinations have been
done using a single electron temperature (T$_e{\rm [O~{III}]}$) for
all the ions.  This approximation introduces errors in the
calculation, especially for the abundances of low ionization ions and,
in turn, of the ICFs.  oxygen and argon are dominated by high
ionization states, while, in the optical, nitrogen can be observed only
in its N$^+$ state; thus the use of T$_e{\rm [N~{II}]}$=T$_e{\rm
[O~{III}]}$ affects the determination of N abundances.  We have tried
to test this effect assuming two different electron temperatures.  The
low ionization temperature is usually derived from the \nii\
I(5755)/I(6583) ratio, which provides T$_e$\nii.  As the \nii\ 5755\AA\
line is not observed in our spectra of M~33, we tried to estimate
whether a relation between temperature from \oiii\ and from the \nii\
line ratio could be established in well studied Galactic PNe.  For
that we have considered the extensive work of KB94.  We have found
that, on average, an approximately linear relation holds between the
\oiii\ and \nii\ temperatures. Using an error-weighted least-squares
fit, we found:
\begin{equation} 
T_e{\rm [N~ II]}=0.23 T_e{\rm [O~III]} + 0.71
\label{rel1} 
\end{equation} 
where T$_e$\oiii\ and T$_e$\nii\, are expressed in 10$^{4}$ K.  
With this relation T$_e$\nii\ can be  crudely estimated for the PNe with known
T$_e$\oiii.  We computed chemical abundances both with
T$_e$\nii=T$_e$\oiii\ and with T$_e$\nii\ from (\ref{rel1}).  With
T$_e{\rm [N~{II}]}<$ T$_e{\rm [O~{III}]}$, the ionic abundance of
N$^+$ is enhanced while the ICF(N) is lowered. The total abundance of
N, however, does not change considerably so that the appreciable
nitrogen deficiency remains.
 
Another effect that we considered in the analysis of the possible N/H
deficiency is the  accuracy of the  flux measurements of the \oii\ doublet at 7325\AA.
The most critical zone of our spectra is indeed the 7200-7400 \AA\ region,
where the sky subtraction is particularly difficult.  The \oii\ flux
is then quite uncertain. This implies possible errors on the O$^+$
abundance, in the ICF(N) and consequently on the N abundance.  The
oxygen abundance is dominated by the \oiii\ flux and the errors on
O$^+$ ionic abundance do not affect it greatly, whereas the ICF(N)
depends on the ratio between the O and O$^+$ abundances (see the
prescriptions by KB94).  We first consider the extreme case in which
the sky subtraction  had been plainly wrong in all three PNe and our mean ionic abundance
of O$^+$ was only one half of the value that follows from Table
3. The N abundances would turn to be comparable with the mean Galactic
PNe N/H value. However, if the error in the sky
subtraction was not so large and systematic the
possibility remains of a real under-abundance in N for PNe of M33.  To
better evaluate this possibility we have compared the behaviour of $\frac{N^+}{O^+}$ 
versus $\frac{N^+}{H}$ for our PNe, which are  disc PNe, 
with that of the Galactic sample  of non-type 1 PNe of KB94.  
Our PNe fall in the lower-left part of the diagram, 
 corresponding to  the lower  $\frac{N^+}{O^+}$ values  of the PNe of  the KB94 sample. 
Our $\frac{N}{H}$, $\frac{N^+}{O^+}$ and also $\frac{N}{O}$ are thus
consistent   with the corresponding values of the lower  $\frac{N^+}{O^+}$ PNe 
of the KB94 Galactic sample.  
If our three PNe would
be representative of the M33 PNe,   their low $\frac{N}{O}$ would indicate 
a poor N enrichment in the atmospheres of the progenitors of the PNe, and 
consequently a N under-abundance.

\begin{figure}[h!]  
\label{nitro} 
\caption{$\frac{N^+}{O^+}$ vs N$^+/H$: empty dots represent non-type 1 PNe from   
KB94 and filled dots the 3 PNe in M33. } 
\resizebox{\hsize}{!}{\includegraphics{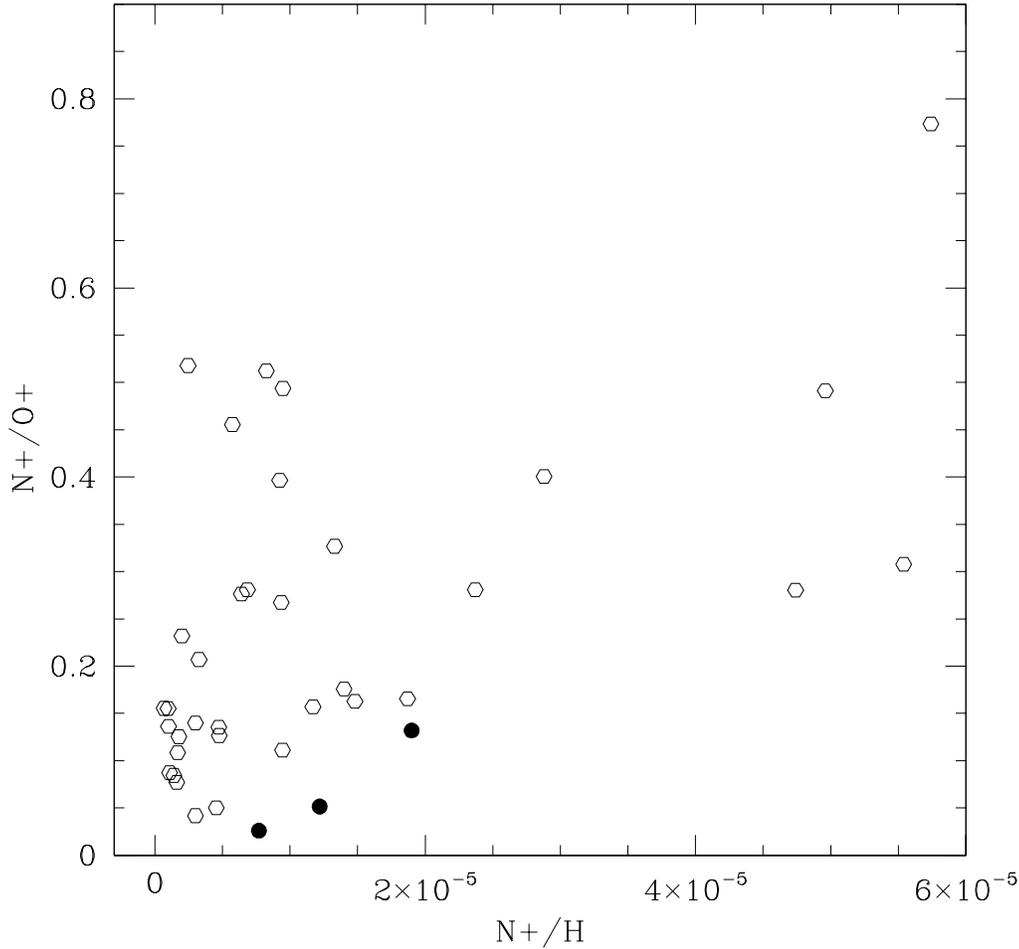}}  
\end{figure}   
   
Obviously, this under-abundance cannot be argued from data of only 3 
PNe  where the T$_e{\rm [N~{II}]}$ is not directly measured, 
but at the same time it cannot be excluded with the 
present data.  
 On the other hand, it seems there is not any evident under-abundance of N in 
\hii\ regions of M33. Their N/O ratio is equal, within
uncertainties, to that of all the galaxies in Table \ref{median} (except 
the determinations by Rubin et al. \cite{rubin}  for
Orion). 
The apparently low  $\frac{N}{O}$ ratio in the studied PNe of 
M33 clearly deserves further investigation.
 
\section{Conclusions} 
 
We have presented optical spectra of a sample of candidate PNe in the
nearby spiral galaxy M~33.  The location of the objects in the
\ha/\sii\ vs \ha/\nii\ diagnostic diagram and their excitation
parameter $R=\frac{[OIII]}{H\alpha+[NII]}$ allowed us to confirm that
26 of the observed 36 candidate PNe with reliable spectra are very
likely genuine PNe, while seven other objects are likely compact
\hii\ regions and another three might be SNR candidates.  None of the 
emission objects that were classified as ``uncertain'' by M00 (because
having a non-negligible continuum emission) seem to be PNe: three are
likely \hii\ regions and two SNRs.  Thus, these spectra confirm the
overall reliability of the criteria adopted by Magrini et
al. (\cite{magrini00},
\cite{magrini01a}, \cite{magrini02}) to select candidate PNe in
galaxies of the Local Group from ground-based narrow-band imaging.
The degree of contamination  found for M33 PNe is approximately 20\%
and 10\% for unresolved \hii\ regions and SNRs, respectively.
 
This work also confirms the potentiality of the use of PNe for the
determination of chemical abundances for the old and intermediate
populations of galaxies, as an alternative to integrated light
spectroscopy.  With a 4~m-class telescope one can
reach the limits of the Local Group and obtain a  direct abundance
determination for PNe, observing the relevant emission lines for
density, temperature and metallicity analysis. We present
a detailed physico-chemical analysis of the three best observed
PNe in M33, finding that He/H, O/H and Ar/H abundances are 
in agreement with the Milky Way PNe values (KB94).  
N/H is unexpectedly low. This might be due to the weakness 
of the \oii\ lines which are strongly involved in the calculation of
the correction factors needed to determine the total Nitrogen
abundance from  single low ionization ions.  On the other hand, the
Galactic behaviour of $\frac{N^+}{H}$ vs. $\frac{N^+}{O^+}$ indicates
that the value of $\frac{N^+}{O^+}$ for the PNe in M33 is comparable
with that of Galactic PNe with low nitrogen abundance.  We cannot then
fully exclude the possibility of a real nitrogen deficiency in PNe of M~33.

\end{document}